\documentclass[aps,prl,twocolumn,superscriptaddress,floatfix]{revtex4}

\usepackage{amsmath,amssymb}
\usepackage{graphicx,multirow}

\begin{document}

\title{Origin of Anomalous Water Permeation through Graphene Oxide Membrane}

\author{Danil W.Boukhvalov}
\email{danil@kias.re.kr}
\affiliation
{Korea Institute for Advanced Study, Seoul 130-722, Korea.}
\author{Mikhail I. Katsnelson}
\email{M.Katsnelson@science.ru.nl}
\affiliation
{Radboud University Nijmegen, Institute for Molecules and Materials, Heyendaalseweg 135, 
6525AJ, Nijmegen, the Netherlands}
\author{Young-Woo Son}
\email{hand@kias.re.kr}
\affiliation
{Korea Institute for Advanced Study, Seoul 130-722, Korea.}

\begin{abstract}
Water inside the low dimensional carbon structures has been considered seriously 
owing to fundamental interest in its flow and structures as well as its practical impact. Recently, 
the anomalous perfect penetration of water through graphene oxide membrane was demonstrated 
although the membrane was impenetrable for other liquids and even gases. The unusual auxetic 
behavior of graphene oxide in the presence of water was also reported. 
Here, based on first-principles calculations, 
we establish atomistic models for hybrid systems composed of water and 
graphene oxides revealing the anomalous water behavior inside the stacked graphene oxides. We 
show that formation of hexagonal ice bilayer in between the flakes as well as melting transition 
of ice at the edges of flakes are crucial to realize the perfect water permeation across the whole 
stacked structures. The distance between adjacent layers that can be controlled either by oxygen 
reduction process or pressure is shown to determine the water flow thus highlighting a unique 
water dynamics in randomly connected two-dimensional spaces.
\end{abstract}

\maketitle
For the last few decades, several studies of the water inside porous carbons, graphite and 
graphene oxide, and carbon nanotubes have provide a plenty of intriguing experimental 
results.~\cite{1,2,3,4,5,6,7,8,9,10,11,12} 
Formation of various kinds of ice in different environment has also been important subject on 
the edge of current physics and chemistry.~\cite{13,14,15,16,17} 
Very recently, the anomalous perfect permeability 
of graphene oxide membrane for water was demonstrated although other liquids and even gases 
cannot penetrate the membrane.~\cite{9} A sudden volume expansion with increasing pressure, {\it i.e.}, 
auxetic material behavior was also reported for graphene oxide in the presence of water.~\cite{10} 
In the work of Nair {\it et al} (Ref. 9), the permeation of water through graphene oxide membrane has been 
explained in terms of anisotropic migrations of water clusters following unoxidized areas within 
graphene oxide sheets~\cite{18,19,20} (see Fig. 1). Brownian (isotropic) motion of the molecules of simple 
gases (He, Ar, N$_2$, H$_2$) does not lead to any noticeable permeation.~\cite{9} 
These observations probably force us to assume some kind of collective motion 
for the anomalous water permeation such as a formation of ice-like structures. 
Interlayer distance in graphene oxide membranes is another 
important factor to control the water flow since the decrease of interlayer distance after reduction 
of graphene oxide makes the membrane impenetrable even for water.~\cite{9} 	

\begin{figure}[b]
\includegraphics[width=1.0\columnwidth]{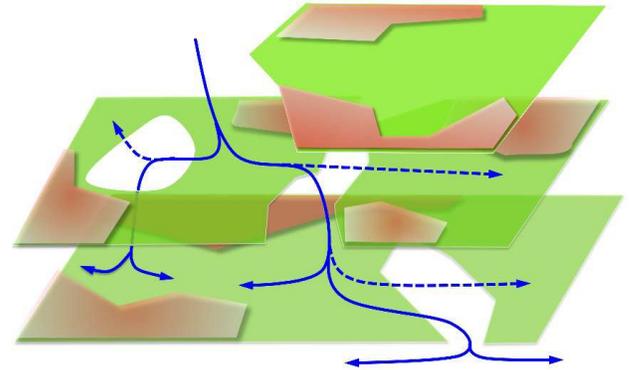}
\caption{
Sketches of water motion through the stacked graphene oxide layers with channels 
(capillaries) in the vicinity of the edges of graphene sheets. The oxidized area is denoted by red 
color and graphene without oxidation by green. Solid (dotted) blue lines are (un)favorable paths 
for water permeations. All edges are assumed to be passivated by hydrophilic edge groups.
}
\end{figure}

To realize liquid or gases permeation throughout randomly stacked two-dimensional flakes, 
three important necessary conditions should be fulfilled. Here we suppose that the single layer 
flake itself is impenetrable for any liquid or gas. First, the capillaries in between two-dimensional 
flakes are required. If small molecules propagate freely in the interlayer space between stacked 
two dimensional materials, they can hardly flow to any specific point, {\it e.g.}, to the edges of flakes. 
Thus, the lateral confinements in a reduced area are particularly helpful to quench the random 
motion of particles. In graphene oxide membrane, the unoxidized part plays as the capillary~\cite{9,18,19,20} 
(Fig. 1). Second, anisotropic energy barriers for migration of liquids or gases in the capillary are 
necessary. Even though some space in between the layers is prepared through the formation of 
capillaries, the isotropic molecular motion inside them will result in a low probability of 
directional flow from one capillary to the other within the interlayer space. Third, assuming that 
they all manage to reach the edge of the flakes, those free molecules will continue to move or 
wandering around in the same interlayer space rather than jump to another interlayer space, thus 
impossible to penetrate across the whole membrane (Fig. 1). So, the energy gain at the passing of 
molecules from one interlayer space to the other is required for a liquid to flow across the 
stacked flakes. For the case of water, according to the recent experiments,~\cite{9} it seems to be a 
collective water flow from one interlayer space to the other but its molecular mechanisms and 
the strong dependence of the effect on the interlayer distance have not been clarified yet.

A systematic investigation of interaction between graphene oxide and water under pressure 
demonstrates a crucial role of the interlayer distances.~\cite{10,11,12} Anomalous negative volumetric 
compressibility in the system has been observed when the interlayer distance in graphene oxide 
was about 9~\AA.~\cite{10} Deviation from this optimal distance makes the system behave as normal 
materials under pressure.~\cite{10} Such a strong dependence on the interlayer distance distinguishes the 
confined water characteristics from those in other carbon nanomaterials. Another key point in 
this two-dimensional system is the possibility of formation of a highly symmetric ice monolayer 
(see for illustration Fig. 2a) over a perfect graphene sheet.~\cite{21} Observation of structural anomalies 
in graphene/alcohols mixes~\cite{22,23} and its absence in graphene/acetone~\cite{24} mix deserve
further investigation on their atomistic structures and flow dynamics.
Despite of many experimental and theoretical works on hybrid systems composed of graphene (oxides) and water, 
a comprehensive picture which might explain the relationship between the interlayer distance 
and structural properties of hexagonal ice formed between graphene layers is still lacking, which 
is also important to understand migration of the ice layers across multiply stacked graphene 
(oxide) layers.

All calculations were performed based on first-principles pseudopotential calculation 
methods~\cite{25} with the generalized gradient approximation and with spin-polarization.~\cite{26} The 
wavefunctions were expanded with a double-$\zeta$ plus polarization basis of localized orbitals for 
carbon and oxygen, and a double-$\zeta$ basis for hydrogen. Optimization of the force and total 
energy was performed within 0.04 eV/\AA~and 1 meV, respectively. For the modeling of graphene 
layers we used a rectangular-like unit cell with 20 carbon atoms in each carbon layer and four 
water molecules in each layer of ice with periodic boundary conditions (see Fig. 2). For the 
checking the role of the supercell size for the energy costs of water migration over hydrophilic 
edges (Fig. 3) we multiply twice the supercell size in direction normal to the breaks of graphene 
sheets and find that the energetic hierarchies do not change. An energy mesh cut-off of 300 Ry 
and a k-point mesh of 8$\times$8$\times$4 are used for our large supercell geometries. We note that our 
method and modeling geometries were successfully used for description of graphene-water 
interfaces~\cite{27,28} and modeling of graphene oxide atomic structure.~\cite{29,30} 
For the additional check the 
calculations of the values of energy barriers with employment of LDA functionals~\cite{31} have been 
employed. Binding energy of the water molecule was calculated as 
$E_\textrm{bind} = (E_\textrm{C+ice}-(E_\textrm{C} + NE_\textrm{W}))/N$, 
where $E_\textrm{W}$ and $N$ are the energy and number of water molecule in the empty box (in 
gaseous phase) respectively, $E_\textrm{C+ice}$ the total energy of graphene or graphite with the hexagonal ice 
structure between carbon layers, and $E_\textrm{C}$ the total energy of the carbon structure before insert of 
ice layers. Binding energy between graphene and ice monolayer is $E_\textrm{bind} = E_\textrm{C+ice}-(E_\textrm{C} + E_\textrm{ice})$, 
where $E_\textrm{ice}$ is the total energy of freestanding hexagonal ice layer.

\begin{figure}[t]
\includegraphics[width=1.0\columnwidth]{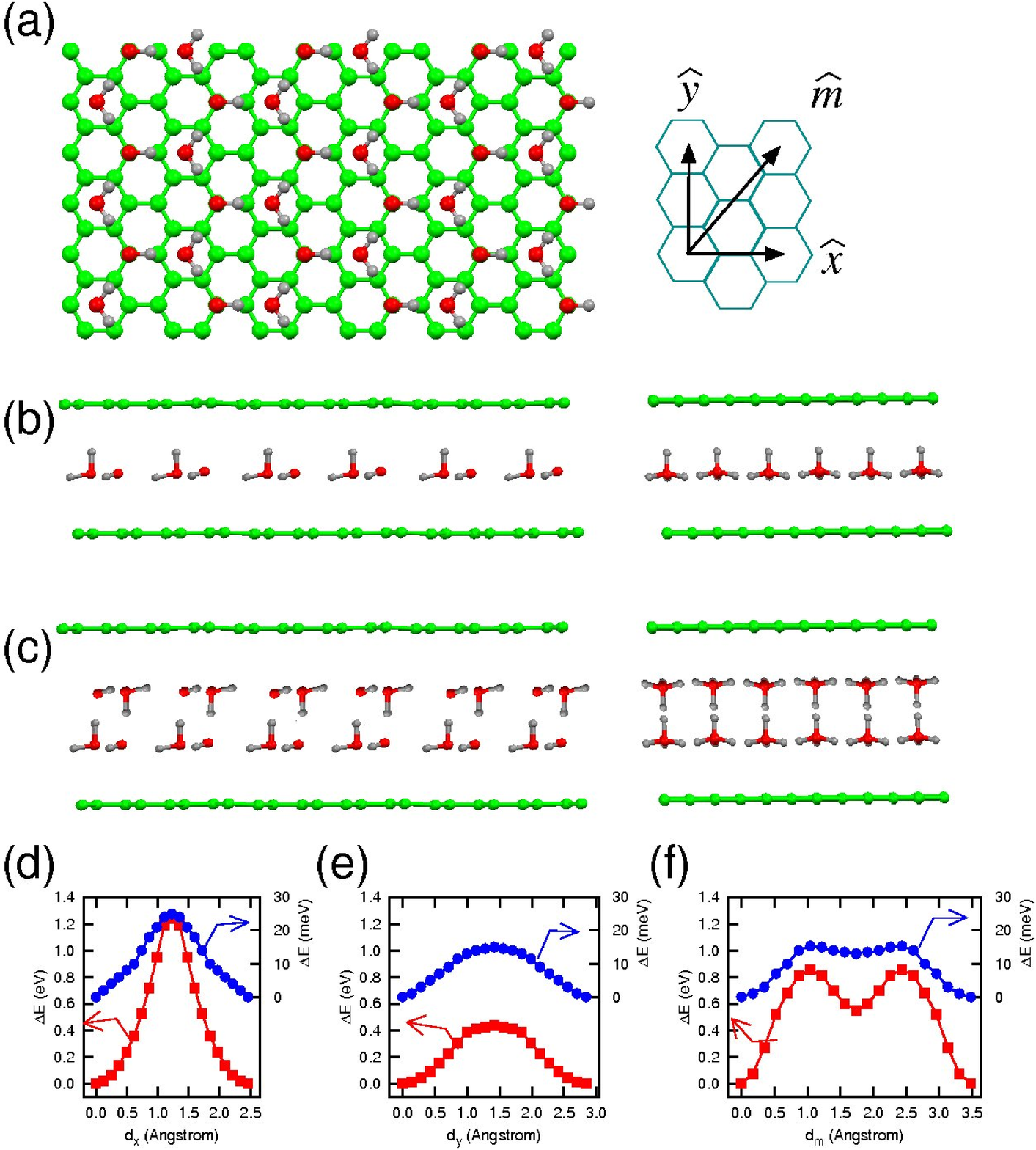}
\caption{
(a) Top view of optimized atomic structure of ice monolayer between the layers of 
graphite. Oxygen, hydrogen, and carbon atoms are denoted by red, grey and green color. The ice 
layer can slide along x-, y- and m-direction. Arrows in the right side denote sliding directions 
and distances. A black arrow along x-direction (armchair direction) has a distance of dx and one 
along y-direction (zigzag) dy. An intermediate direction in between them is denoted by m-
direction and distance of corresponding arrow is dm. (b) and (c), Side view of ice mono- and bi-
layers along y-direction (left panels) and ones along x-direction (right panels). (d), (e), and (f), 
Migration energy barriers for ice sliding along x-, y-, and m-directions. Blue circles are energy 
costs for sliding of ice monolayer shown in (b) and red rectangles for sliding of one ice layer in 
ice bilayer shown in (c).
}
\end{figure}

First, we examine the case of ice monolayer over single-layer graphene (Figs. 2a and 2b). The 
calculated binding energies between water molecules in this structure are quite large ($-1.18$ 
eV/H$_2$O) reflecting very strong hydrogen bonds (the total cohesive energy of the hydrogen bonds 
defined as the difference between the formation energies of water in the liquid and gaseous 
phases is as high as 0.46 eV),~\cite{32} in contrast to the binding energy between graphene and ice 
monolayer ($-50$ meV/H$_2$O). Note that the latter value is still much larger than typical van der 
Waals energies, which are usually less than 10 meV,~\cite{33} due to polarity of water molecule. The 
results obtained demonstrate that graphene is the matrix for the formation of layered hexagonal 
ice structure (Fig. 2a). The hexagonal structure of graphene scaffold provides an initial pattern 
for the ordering of water molecules in an energetically favorable structure with hexagonal 
symmetry and the axes coinciding with the crystallographic axes of graphene. Combination of 
weak ice-graphene interactions and strong internal interactions within the ice layer enables the 
whole ice layer slide over graphene, as was suggested in Ref. 9. 

We also examined the formation of this water monolayer between the two graphene layers, 
between every second layers in graphite, and between every layer in graphite, as well as various 
types of their stacking orders different from ordinary Bernal type one. The interlayer distance 
between graphene layers separated by the ice monolayer is about 6\AA~(5.74 $\sim$ 5.87 \AA), the value 
discussed in the literature10 as a minimal distance in graphene oxide/water systems. This value is 
very close to the interlayer distance in reduced graphene oxide impenetrable for water.~\cite{9} The 
presence of carbon layer from the other side of ice monolayer has negligible contributions (about 
$5\sim10$ meV/H$_2$O for different types of stacking) in changing the binding energy between water 
molecules in ice and between ice and graphene. To model the motion of water monolayer 
between graphene planes we moved the monolayer along different directions and calculated the 
energy relief (Figs. 2d-2f). Energy barriers were calculated as $15\pm5$ meV/H$_2$O along zigzag 
direction, $25\pm5$ meV/H$_2$O along armchair direction, and $17\pm5$ meV/H$_2$O along intermediate 
direction (errorbars $\pm$ corresponds to various types of stacking orders). Values of these energy 
barriers calculated within LDA are 4 meV higher. Thus, the motion of ice monolayer in between 
graphene layers can be concluded as isotropic.  

\begin{figure}[t]
\includegraphics[width=1.0\columnwidth]{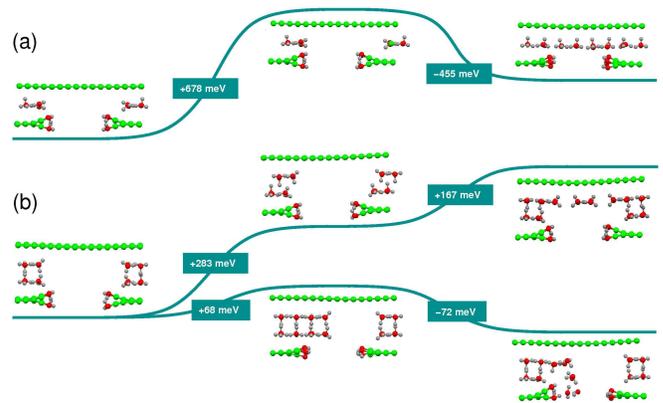}
\caption{
(a) Energy costs (meV/H$_2$O) and optimized atomic structures of migrations of ice 
monolayer along zigzag direction over the hydroxyl group passivated edges in stacked graphene 
oxides. (b) Same diagrams for ice bilayer migration over the edges. The upper migration 
pathway denotes a sliding of top layer in the ice bilayer and the lower one a destruction of ice 
structure (melting) at void space near edges.
}
\end{figure}

Let us have a look now at the interaction between water monolayer and the edge of graphene 
flake. To simulate this, we consider a hybrid system composed of water monolayer and graphene 
double layer; the water monolayer is sandwiched between a perfect top graphene layer and an 
imperfect bottom graphene layer with a hole formed by eliminating several rows of carbon atoms 
along armchair direction. The edges in the bottom layer were passivated by hydroxyl groups 
(Fig. 3a) as a proper way to model hydrophilic edges of graphene oxide sheets.~\cite{34} We calculate 
the energy costs for the migration of water monolayer through the edge, similar to the way how it 
was done for the migration relief across the layer spaces in the bulk. The computational results 
(Fig. 3a) show that, due to interactions of water molecules with hydroxyl groups, the ice 
structure was strongly distorted and the height of the barriers was increased in the two orders of 
magnitude in comparison with the bulk case. This means that, when reaching the edge, it is much 
more energetically favorable for the water monolayer to continue to slide along the carbon sheet 
than to pass through the hole within graphene oxide flakes. Since the distance between graphene 
layers with ice mono-layer (Fig. 2b) corresponds to the typical one for reduced graphene oxide 
we have explained in this way why this material is impenetrable for water, in agreement with the 
experimental observations.~\cite{9}  

For larger interlayer distances, formation of water bilayer between graphene oxide sheets is 
possible. So, we switch to the modeling of this case. We have performed the calculations of 
different types of stacking of carbon sheets in double layer graphene and graphite and various 
mutual orientations of the ice layers and found that the structures shown in Fig. 2c are the most 
energetically favorable for ice structures between graphene layers as well as between layers in 
graphite.  The binding energy between water molecules in the ice bilayer was calculated as $-1.63$
eV/H$_2$O, that is, a hexagonal ice bilayer is more energetically favorable than monolayer due to 
formation of additional hydrogen bond per each water molecule with a help of the enough space 
for its formation. Similar to the case of water monolayer, the axes of the formed bilayer 
hexagonal ice coincide with the crystallographic axes of graphene. The equilibrium distance 
between graphene layers separated by the ice bilayer is within the range $8.43\sim8.59$\AA~for 
different types of stacking corresponding to the range for the interlayer distances in stacked 
graphene oxide allowing water permeation ($7\sim10$\AA).~\cite{9,13} 
Calculations of the energy barriers for 
the shift of one of ice layer over another (see Figs. 2d-2f) demonstrate much higher values of the 
migration barriers than for ice monolayer case considered before and, more importantly and less 
trivial, a significant anisotropy of the migration energy relief. Calculations of the same barriers 
within LDA leads increasing of the values about 60 meV that corresponding with overestimation 
of bonds strengths within this method.  We find that the migration of one ice layer along zigzag 
direction is the most favorable. The cause of colossal energy barriers for the sliding in armchair 
direction is the passing of hydrogen atoms of water molecules from one layer in the vicinity 
(about 0.1\AA) of the hydrogen atoms from the other layer (Fig. 2c). For the case of migration 
along zigzag direction, the distance between hydrogen atoms is the largest and therefore the 
energy barriers are minimal. The energy cost for this case mainly originates from the hydrogen 
bonds breaking between ice layers. Considering both the energy required for the water 
evaporation at the room temperature (0.46 eV/H$_2$O)~\cite{32} and additional pressure induced by 
capillaries in the systems, we expect that the barrier can be easily overcome in the experimental 
conditions.~\cite{9,10} Thus, we can conclude here that the interlayer distance in the stacked graphene 
oxide flakes is quite optimal allowing formation of ice bilayer in the interlayer space and its 
anisotropic water flowing, otherwise only random Brownian movements of water occur.

We have considered also energetics of motion of the second ice layer over the edge of 
graphene oxide flake, similar to the case of ice monolayer discussed above. It turned out that the 
migration of the second ice layer through the edge along zigzag direction (Fig. 3b) is 
energetically more favorable (by $\sim$100 meV/H$_2$O) than its further gliding along the first ice layer 
in the same plane. The reason is that the distortions of ice structure due to interaction with 
hydroxyl groups associated with the edge decrease the interlayer interaction in ice. Further, 
alternative processes are either migration of the ice layer in the plane or a shift of water 
molecules from top layer to the bottom one with further destruction (melting) of ice (Fig. 3b). 
Our calculations show that the last process is the most energetically favorable. Thus, edges of 
graphene oxide passivated by hydrophilic groups stimulate a destruction of ice bilayer (but not 
monolayer) with penetration of water molecules to the next void where they form another ice 
bilayer. This is the model of the process shown schematically in Fig. 1. We believe therefore that 
the formation of ice bilayer in unreduced graphene oxide is the crucial element in anomalous 
penetration of water through graphene oxide paper.~\cite{9}

\begin{figure}[t]
\includegraphics[width=1.0\columnwidth]{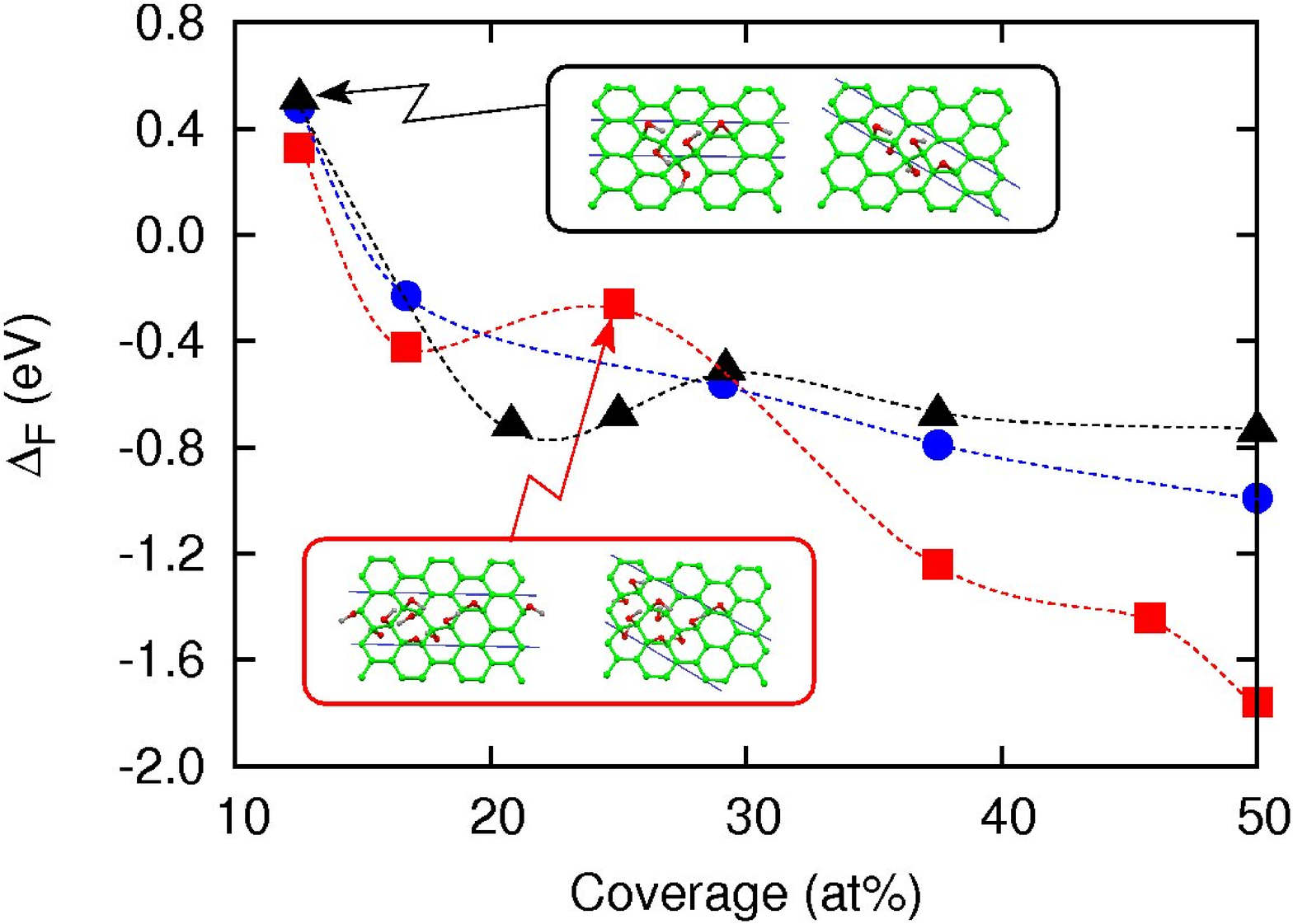}
\caption{$\Delta_F$ (in electron volt) is a total energy difference between the oxidized graphene 
structures with capillary formation along zigzag and armchair directions as function of oxygen 
coverage (percentage of carbon atoms covalently bonded to oxygen atoms). Red rectangles (blue 
circles) are for $\Delta_F$ when a number of epoxy groups is smaller (larger) than one of hydroxyl 
groups. Black triangles are for $\Delta_F$ only with hydroxyl groups. Inset inside a black (red) box 
shows optimized atomic structures with 12.5 at\% (25 at\%) oxygen coverage. In each inset, a left 
(right) panel corresponds to atomic structures showing armchair (zigzag) direction capillary 
formation. Dotted lines are guides for eyes.
}
\end{figure}

The anisotropy of the migration of water in the case of ice bilayer can explain why water 
penetrates throughout graphene oxide multilayers only with interlayer distances $7\sim10$\AA.~\cite{9} We 
have to consider, as the next step, the role of capillaries formed by reduced graphene regions in 
unreduced graphene oxide. We have made a series of calculations for coexisting reduced and 
unreduced areas. One can easily see that in the typical graphene oxides (about 50\% of carbon 
atoms are connected with oxygen or hydroxyl groups) capillaries from unoxidezed area can 
connect each other across all stacking layers regardless of random oxidation area distributions in 
each layer. Next, we have model the step-by-step oxidation of graphene sheet with formation of 
graphene oxide for the various coverage of hydroxyl and epoxy groups. We have started with the 
pair of hydroxyl groups (as more stable) and added the next pair of hydroxyl groups or epoxy 
groups in according to the previously discussed principles of graphene oxide functionalization~\cite{29,30} 
with the tendency to form oxidized patterns along zigzag or armchair directions (Fig. 4). The 
computational results demonstrate that for the small amount of epoxy and hydroxyl groups 
(below 15\% that corresponds to the case of reduced graphene oxide) the formation of stripes 
along armchair direction is more energetically favorable.  With the increase of the number of 
functional groups, however, the preferable direction of stripe-like oxidized areas switches to 
zigzag direction (Fig. 4). The continuous web of capillaries oriented along zig-zag direction 
plays probably a crucial role in water permeation. Independence of the favorability of patterns 
oriented along zig-zag direction from the ratio of epoxy and hydroxyl groups suggest for the 
preference of these capillaries in all types of unreduced graphene oxides with different exact 
atomic structures.

To be complete, we have performed the same modeling for the case of ice trilayer between 
graphene sheets. The calculated equilibrium distance between graphene layers is about 
$11.38\sim11.46$\AA~for different types of graphene stacking.  Following the same calculation 
procedures for ice mono- and bi-layer in between graphene layers, we find that the ice trilayer 
energetics are quite similar to those of monolayer. These calculations therefore can explain the 
reason why the auxetic behavior of graphene oxide realizes in the very specific condition~\cite{10}
because the only optimal interlayer distance for ice bilayer formations can allow water filling 
through graphene oxide against the pressure. All considerations hitherto thus support that the 
formation ice bilayer inside graphene oxide membranes and their melting behaviors owing to 
their optimal interlayer distance are the crucial reasons why they allow perfect permeation for 
waters not for others~\cite{9} and why they exhibit the auxetic behaviors of graphene oxide inside 
waters.~\cite{10}

\acknowledgments
M. I. K. acknowledges support from FOM  (the Netherlands). Y.-W. S. was supported by the 
NRF of Korea grant funded by the MEST (CASE, 2011-0031640 and QMMRC, No. R11-2008-053-01002-0). 
Computations were supported by the CAC of KIAS.

\end{document}